\documentclass[14pt,a4paper]{article}

\usepackage{amsfonts,amssymb,amsmath,amsthm}
\usepackage{xcolor}
\usepackage[colorlinks=true,linkcolor=black,citecolor=teal,urlcolor=blue,filecolor=black]{hyperref}
\usepackage{footnote}
\usepackage{authblk}
\usepackage{float}
\usepackage{physics}
\usepackage{cite}
\usepackage[margin=2cm]{geometry}

\usepackage{tikz}

\usepackage{graphicx}
\usepackage{dcolumn}
\usepackage{bm}
\usepackage[english]{babel}

\let\arcsinh\relax
\let\arccosh\relax
\let\arctanh\relax
\let\sgn\relax

\DeclareMathOperator{\arcsinh}{arcsinh}
\DeclareMathOperator{\arccosh}{arccosh}
\DeclareMathOperator{\arctanh}{arctanh}

\DeclareMathOperator{\sgn}{sgn}

\newcommand{\secref}[1]{Section~\ref{#1}}

\newcommand{\ur}{(u/a)}
\newcommand{\ua}{(u/a)}
\newcommand{\fermi}{\text{v}_F} 

\begin{document}

\title{Classical traversability of a carbon BTZ wormhole}

\author[1]{Valerie Tynianska}

\author[2,1]{Alfredo Iorio}

\affil[1]{Institute of Particle and Nuclear Physics, Faculty of Mathematics and Physics \\
Charles University, V Hole\v{s}ovi\v{c}k\'ach 2, 18000 Prague, Czech Republic}
\affil[2]{Istituto Nazionale di Fisica Nucleare, Viale Berti Pichat 6/2, 40126 Bologna, Italy}


\maketitle

\begin{abstract}
We investigate the classical traversability of an analogue spatial wormhole geometry realized on a two-dimensional curved carbon membrane. The background is obtained by implementing a reflection-symmetric doubling technique on the exterior spatial sector of a non-rotating Ba\~nados-Teitelboim-Zanelli black hole, effectively replacing the event horizon with a smooth traversable throat. Within this framework, massless Dirac quasiparticles propagate as spinorial light rays whose kinematics are governed exclusively by the metric curvature, forcing them to track the null geodesics of the ultrastatic spacetime. By evaluating the exact geodesic equations, we demonstrate that classical traversability is dictated by a sharp binary threshold in the conserved angular momentum, where the minimum throat radius acts as a centrifugal potential barrier that separates traversing paths from reflecting ones. We show that near-critical trajectories experience a logarithmic divergence in their angular evolution, causing the quasiparticles to execute multiple spatial windings around the throat before escaping. These classical ray-tracing configurations establish the necessary geometric backbones for future mesoscopic investigations.
\end{abstract}



\section{Introduction} \label{intro}

The study of wormhole (WH) solutions in general relativity (GR) and quantum gravity (QG) has recently undergone a significant renaissance \cite{wittenPrincetonTalk}. Historically initiated by the seminal Einstein-Rosen bridge \cite{einstein1935} and later refined by Morris and Thorne into the paradigm of traversable Lorentzian wormholes \cite{morris1988}, these geometric shortcuts across spacetime originally required energy-condition-violating matter fields to remain open.

While probing cosmological wormholes remains beyond current observational capabilities, condensed-matter analogues offer a promising laboratory framework for simulating aspects of these geometries on a tabletop scale. The pioneering work of Gonz\'alez and Herrero \cite{gonzalez2009} investigated a simple wormhole geometry consisting of two graphene sheets connected by a short nanotube, exploring the electronic excitation modes associated with the junctions. This work established a connection between geometry, topology, and electronic structure, which was subsequently explored in numerous studies addressing a variety of physical effects, including those of external magnetic fields \cite{alencar2021, cavalcante2022, dogan2023, garciafurtado2019, garciafurtado2025}. More recently, the theoretical framework has been extended to more sophisticated wormhole throat geometries, including structurally asymmetric catenoid wormholes \cite{gorza2026asymmetric, naderi2024}. Collectively, these developments highlight the potential of carbon-based geometries as analogue systems for investigating quantum phenomena in curved backgrounds.

In this work we focus on the classical traversability through these geometric bridges, a transport problem that remains open in analogue gravity. Specifically, we frame our study by focusing on the constant-time spatial slice of the Ba\~nados-Teitelboim-Zanelli (BTZ) black hole geometry realized in graphene. This choice is motivated by earlier investigations of hyperbolic structures of constant negative Gaussian curvature, ${\cal K} = - 1/a^2$, within the context of quantum field theory (QFT) in curved graphene spacetimes \cite{iorio2014,iorio2012}, see also the reviews \cite{iorio2015,universegraphene}. Here, we shift the perspective from frameworks based primarily on the local Weyl symmetry \cite{lor} of the emergent Dirac theory \cite{i2,weylgraphene} to an approach based on taking constant-time slices of the BTZ black hole spacetime, as required to model traversable wormhole geometries.

To model this transport problem, we rely on a specific hierarchy of scales. First, we enforce the long-wavelength approximation with respect to the lattice spacing ($\lambda_e \gg \ell$), ensuring that the charge carriers behave as continuous massless Dirac fermions near the charge neutrality point \cite{PacoReview2009}. Furthermore, by requiring the excitation wavelength to be small compared to the geometric parameters associated with the curvature, $a$, and the throat radius, $c$, such that $a,c \gg \lambda_e \gg \ell$, we access a semiclassical eikonal regime characterized by large transverse quantum numbers \cite{rubinow1963}.

Under these conditions, the analysis simplifies via systematic geometric constraints. At the leading order of the short-wavelength expansion, the algebraic consistency of the Dirac equation requires the matrix operator to be non-invertible to sustain a non-zero current. This constraint yields a scalar Hamilton-Jacobi equation, which decouples the spinorial path determination from the physical trajectories. Within this eikonal framework, both the intrinsic spin connection and potential lattice contorsion contributions are structurally confined to the sub-leading orders of the asymptotic expansion. Consequently, at the dominant order, the quasiparticles decouple from these degrees of freedom and follow the pure metric geodesics of the background geometry.

Finally, within a $(2+1)$-dimensional framework, the transverse shear and vorticity of a null geodesic bundle vanish identically due to the one-dimensional nature of the transverse subspace. The general Raychaudhuri equation thus simplifies structurally from the outset. When applied to our specific hyperbolic geometry, the remaining curvature terms prevent gravitational focusing, causing the relation to collapse into a regular Jacobi equation. The absence of curvature-induced caustics (focal points) for beams that are not strongly converging ensures that independent, non-intersecting geodesic trajectories are sufficient to track the global transport of charge carriers across the throat.

Using this geometric framework, we investigate in detail the kinematics of both traversing and reflecting geodesics. We show that classical traversability is governed by the interplay between the minimum throat radius $c$ and the conserved classical angular momentum $|J|$ of excitations arising from the first integral of the geodesic equations. Trajectories with conserved angular momentum $|J|<c$ traverse the throat, whereas those with $|J|>c$ encounter an effective centrifugal barrier and are reflected. The critical case $|J|=c$ corresponds to trajectories that asymptotically approach the throat without crossing it.

Crucially, we show that near-critical trajectories, $|J| \simeq c$, experience a logarithmic divergence in their angular evolution and an increasing number of windings around the throat. In the classical ballistic limit, these multiple windings introduce a kinematic time delay without altering the stationary current, establishing the structural backbone for future quantum transport studies where these orbiting states are expected to trigger quantum interference resonances.

The structure of the paper is as follows. In Section 2, to set up the logic and the notation, we review the use of graphene as a tabletop quantum field theory in curved space under the Weyl-symmetry perspective. In Section 3, we analyze the spatial slices of the BTZ spacetime and identify their exact geometric mappings onto the three fundamental pseudospheres. Section 4 formalizes the construction of the effective traversable wormhole metric. Section 5 outlines our semiclassical strategy based on the (2+1)-dimensional Raychaudhuri equation. Section 6 presents the full classification and exact integration of light-like geodesics on the hyperbolic geometry. Section 7 introduces the gravitational lensing setup and calculates the deflection angle, while Section 8 addresses the ray-limit transmission threshold. Section 9 contains our conclusions and future outlook. Finally, the technical derivations and geometric details are organized in four Appendices. Appendix A reviews the differential geometry of the three proper pseudospheres; Appendix B details the operational necessity of conformal flatness; Appendix C provides the exact geodesic integration for the elliptic case; and Appendix D contains the rigorous semiclassical WKB expansion and spin-torsion decoupling constraints.

\section{Graphene as a laboratory for gravity: the Weyl-symmetry approach}\label{StatOfArt}

Theoretical investigations, conducted prior to the experimental isolation of graphene, established that the low-energy behavior of conduction electrons near the Fermi energy can be mapped onto a special relativistic field theory in flat space \cite{semenoff1984}. This structural analogy has been evaluated since the isolation of the material, serving as a framework for high-energy physics simulations on tabletop systems (see, e.g., \cite{KATSNELSON2007}).

Given that the structural deformations of the carbon membrane modify the electronic properties of graphene, research has focused on the effects of curvature, both extrinsic and intrinsic \cite{PacoReview2009}. This line of investigation follows the approach in Ref.~\cite{VozmedianoPRL1992}, where methods based on the Dirac equation in curved space were employed to reproduce the electronic spectrum of giant fullerene molecules. Within this framework, the mathematical methods of QFT in curved spacetimes are used to describe the corresponding condensed matter phenomenology.

An alternative perspective was introduced with the implementation of the local Weyl-gauge symmetry approach, for massless Dirac particles in curved backgrounds \cite{lor}. This formulation allows the investigation of deformed graphene sheets as physical analogues of curved spacetimes, connecting condensed matter systems with models studied in gravitational and high-energy physics theory \cite{weylgraphene}. This approach represents an active line of research with several theoretical developments \cite{iorio2012,iorio2014,iorio2018generalized,iorio2018anti,ip5,iorio2020generalized,ciappina2020torsion,threelayers,NCthreelayers,pais2026valley} and numerical validations \cite{LobLab_2016,Klaver2025} aimed at experimental realizations (see also the reviews \cite{iorio2015,universegraphene}). In particular, previous works \cite{iorio2012,iorio2014} focused on identifying distinct electronic signatures of the Hawking-Unruh effect within the local density of states (LDOS).

The local Weyl symmetry constitutes a central element in this framework. The physical line element describing the deformed carbon membrane is typically chosen to be of the ultrastatic form
\begin{equation}\label{mainmetric}
    ds^2_{\text{phys}} = g^{\rm graphene}_{\mu \nu} dx^\mu dx^\nu = dt^2 - g_{i j} (x,y) dx^i dx^j \,,
\end{equation}
where only the spatial sector is non-trivial. While this highly restrictive choice represents a simplified physical scenario, where a curved spatial background is generated by deforming the membrane in a steady configuration, the absence of time-dependent components explicitly prevents the metric $g^{\rm graphene}_{\mu \nu}$ from hosting a standard Killing horizon (KH).

To overcome this constraint and evaluate horizon-induced thermodynamic phenomena, the local Weyl symmetry of the massless Dirac action in $d=3$ dimensions is exploited \cite{lor,weylgraphene}. Under a local Weyl gauge scaling, the metric and the spinor field transform according to:
\begin{align}\label{WeylTrafos}
    g^{\rm graphene}_{\mu \nu} &\rightarrow g^{\rm Weyl}_{\mu \nu} = e^{2 \sigma(x,y)} \, g^{\rm graphene}_{\mu \nu}, \\
    \psi &\rightarrow \psi^{\rm Weyl} = e^{- \sigma(x,y)} \, \psi.
\end{align}
The conformally related metric $g^{\rm Weyl}_{\mu \nu}$ acquires nontrivial time components via the space-dependent conformal factor $e^{2 \sigma(x,y)}$. Consequently, if one identifies a conformal transformation relating the physical metric \eqref{mainmetric} to a target metric $g^{\rm Weyl}_{\mu \nu}$ that possesses a standard KH, the physical system exhibits a \textit{conformal Killing horizon} (CKH) at the location of the target horizon.

Furthermore, we enforce \textit{conformal flatness}, which in a $(2+1)$-dimensional manifold is equivalent to the vanishing of the Cotton tensor. For the synchronous-like ansatz \eqref{mainmetric}, this condition uniquely requires the spatial slice to be a Riemannian surface of \textit{constant intrinsic Gaussian curvature} $\cal K$ \cite{i2}. Since a positive curvature sphere (${\cal K} = + 1/a^2$) does not lead to standard event horizons \cite{iorio2014}, the three classic pseudospheres of constant negative Gaussian curvature (${\cal K} = - 1/a^2$) are chosen. Conformal flatness acts as an algebraic condition that accommodates a flat laboratory time evolution while selecting highly symmetric spatial backgrounds, enabling non-perturbative evaluations of quantum correlators (see Appendix B for details).

By exploiting the local Weyl covariance of the massless Dirac action, the two-point spinor Wightman function on the physical pseudospherical background, $S^{\rm phys}(q_1, q_2)$, maps directly onto the reference horizon correlator, $S^{\rm ref}(q_1, q_2)$, according to the rule of conformal triviality \cite{birrell1984,iorio2014}:
\begin{equation}
    S^{\rm phys}(q_1, q_2) = \varphi^{-1} (q_1) \varphi^{-1} (q_2) S^{\rm ref}(q_1, q_2),
\end{equation}
where $\varphi = e^{\sigma}$ is the conformal factor and $q^\mu = (t, u, v)$ represents the inner coordinate frame. When evaluated at spatial coincidence ($\vec{q}_1 = \vec{q}_2 \equiv \vec{q}$), the geometric impact of the conformal factor does not warp the intrinsic KMS periodicity of the reference thermal bath. Instead, it factorizes as a space-dependent multiplicative scaling of the entire two-point function. The physical LDOS, $\rho^{\rm phys}(\omega, \vec{q})$, inherits this factorization under the Fourier transform with respect to the laboratory time $t \to \omega$:
\begin{equation}
    \rho^{\rm phys}(\omega, \vec{q}) = \varphi^{-2}(\vec{q}) \rho^{\rm ref}(\omega, \vec{q}),
\end{equation}
where $\rho^{\rm ref}(\omega, \vec{q})$ carries the unrescaled thermal signature of the reference horizon.

Depending on the chosen negative-curvature surface/membrane, the physical manifold maps conformally to a specific spacetime. Of those, the spacetimes associated to the three pseudospheres, play a distinguished role:
\begin{enumerate}
    \item \textit{The Beltrami Pseudosphere $\rightarrow$ Rindler Spacetime.} \\
    When $g_{ij}$ is the Beltrami pseudosphere (BEL), the reference metric corresponds to a flat Rindler wedge, where the boundary of the surface (the ``Hilbert horizon'') serves as the event horizon. The reference LDOS $\rho^{\text{ref}}$ yields an Unruh distribution at the standard temperature $T_U = \hbar \alpha / (2\pi v_F k_B)$, where the acceleration parameter is geometrically fixed by the pseudosphere radius as $\alpha = v_F^2 / a$.

    \item \textit{The Elliptic Pseudosphere $\rightarrow$ de Sitter Spacetime.} \\
    The elliptic pseudosphere (ELL) maps the system conformally onto the open, hyperbolically sliced patch of de Sitter space bounded by a cosmological horizon. The physical manifold exhibits a CKH tracking a conformal manifestation of the Gibbons-Hawking effect governed by $T_{\text{dS}}$.

    \item \textit{The Hyperbolic Pseudosphere $\rightarrow$ BTZ Black Hole and Wormholes.} \\
    When $g_{ij}$ corresponds to the hyperbolic pseudosphere (HYP), the system maps conformally onto the exterior region of the $(2+1)$-dimensional Ba\~nados-Teitelboim-Zanelli (BTZ) black hole metric \cite{casals2019quantum}. By applying the reflection-symmetric doubling technique, this configuration describes the geometry of a traversable analog wormhole.
\end{enumerate}

This framework provides the methods and mathematical instruments to address the cases investigated in the present paper.

\section{The spatial slice of the BTZ spacetime as a pseudosphere in $\mathbb{R}^3$}
\label{BTZspatial}

\begin{figure}[t]
    \centering
    \includegraphics[width=0.7\linewidth]{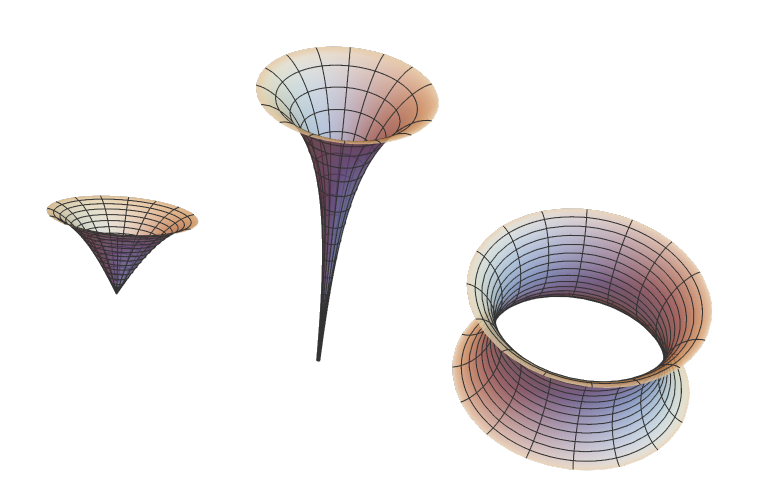}
\caption{The three pseudospheres: (left to right) elliptic (ELL), Beltrami (BEL), hyperbolic (HYP).}
    \label{fig:3 pseudosp ELLBELLHYP}
\end{figure}

The BTZ metric \cite{BTZ1992} for a non-rotating BH (vanishing BTZ angular momentum) reads
\begin{equation}\label{BTZ-metric}
    ds_{BTZ}^2=\left(\frac{\mathcal{R}^2}{l^2}-M\right)d\mathcal{T}^2- \left(\frac{\mathcal{R}^2}{l^2}-M\right)^{-1} d\mathcal{R}^2 -\mathcal{R}^2 d{\mathcal{V}}^2 \,,
\end{equation}
where the cosmological constant $\Lambda = - \, l^{-2}$ is negative. The horizon is reached, when
\begin{equation}\label{defRhor}
    \mathcal{R} = l \, \sqrt{M} \equiv \mathcal{R}_{H} \ .
\end{equation}
Fixing the time coordinate at $\mathcal{T}=\mathrm{const}$ in the BTZ metric \eqref{BTZ-metric}, we obtain the spatial slice
\begin{equation} \label{space-part-BTZ}
    dl^2_{BTZ} = \left(\frac{\mathcal{R}^2}{l^2}-M\right)^{-1} d\mathcal{R}^2 + \mathcal{R}^2 \, d{\mathcal{V}}^2 \,.
\end{equation}
Introducing the proper radial coordinate $u$ through
\begin{equation}\label{du-dR-BTZ}
    du=\frac{d\mathcal{R}}{\sqrt{\frac{\mathcal{R}^2}{l^2}-M}}
\end{equation}
brings the metric \eqref{space-part-BTZ} into the canonical form of a rotationally symmetric two-dimensional metric
\begin{equation}\label{BTZ-spatial-canonical}
    dl^2_{BTZ}=du^2+\mathcal{R}^2(u) \, d\mathcal{V}^2 \,,
\end{equation}
which coincides with the metric induced on a surface of revolution (motivated in earlier works on the BTZ spectrum, see e.g. \cite{brill2000black,brill20002+,brill1996geometry}).

A surface of revolution embedded in $\mathbb{R}^3$ can be parametrized as
\begin{equation}\label{embedding-surf-of-revolution}
    \vec{\sigma}(u,v)=(R(u)\cos v, R(u)\sin v, z(u)) \,,
\end{equation}
where the parametrization is chosen such that
\begin{equation}\label{arc-length-parametrization}
    z'^2(u)+R'^2(u)=1 \,.
\end{equation}
Then, the induced metric takes the form
\begin{equation}\label{surface-of-revolution-metric}
    dl^2 = du^2+R^2(u) \, dv^2 \,,
\end{equation}
with Gaussian curvature ${\cal K} = -\frac{R''(u)}{R(u)}$.

The canonical form \eqref{BTZ-spatial-canonical} suggests that the spatial slice of the BTZ spacetime can be locally represented as a surface of revolution embedded in $\mathbb{R}^3$. Comparing \eqref{BTZ-spatial-canonical} and \eqref{surface-of-revolution-metric}, we identify
\begin{equation}\label{identification}
    R(u) = \mathcal{R}(u) \,, \ v = \mathcal{V} \,.
\end{equation}

With this identification, the radial coordinate $\mathcal{R}(u)$ of the BTZ spacetime represents the profile function $R(u)$ of the surface of revolution, and Eq. \eqref{du-dR-BTZ} becomes
\begin{equation}\label{surface-classification-eqn}
    \frac{dR(u)}{du} = \frac{1}{l}\sqrt{R^2-l^2 M} \,.
\end{equation}

The BTZ spacetime admits a continuation to negative values of the BTZ mass parameter $M$. Since the BTZ solution is obtained by discrete identifications of the global AdS$_3$ spacetime, different values of $M$ describe distinct spacetime configurations \cite{PhysRevD.48.1506,casals2019quantum}. For example, negative values of the mass parameter ($-1<M<0$) correspond to the conical-defect sector of the BTZ spectrum, commonly interpreted as a point particle located at the origin of an otherwise empty AdS$_3$ spacetime, with the particle's mass acting as a localized source of the conical deficit \cite{Lukes2026}. The black-hole part of the BTZ spectrum begins with the special case of the zero-mass black hole ($M=0$) and continues to positive values of the mass parameter ($M>0$).

This means that, following the BTZ spectrum in all generality, the mass parameter $M$ introduces three cases: $M<0$, $M=0$, and $M>0$, which we need to consider when solving Eq.~\eqref{surface-classification-eqn} for the profile function $R(u)$. Integrating \eqref{surface-classification-eqn}, we obtain
\begin{align}
    M<0: \, u &= \int \frac{l\,dR}{\sqrt{R^2-l^2M}}, & R(u)&=l\,\sqrt{-M}\sinh(u/l)\,, \label{BTZ-profile-ELL} \\
    M=0: \, u &= \int \frac{l\,dR}{R}, &  R(u)&=\beta e^{u/l} \,, \label{BTZ-profile-BEL} \\
    M>0: \, u &= \int \frac{l\,dR}{\sqrt{R^2-l^2M}}, & R(u)&=l\,\sqrt{M}\cosh (u/l) \, \label{BTZ-profile-HYP} .
\end{align}

We choose the integration constant $u_0=0$, as it just fixes the origin of the $u$-coordinate for the cases $M<0$ and $M>0$, whereas in the case $M=0$ the arbitrary integration constant $\beta$ remains in the solution.

Solutions~\eqref{BTZ-profile-ELL}--\eqref{BTZ-profile-HYP} coincide with the profile functions~\eqref{ELL}--\eqref{HYP} defining the three pseudospheres: ELL, BEL, and HYP (see Appendix \ref{appendixPseudospheres} for details) pictured in Fig.~\ref{fig:3 pseudosp ELLBELLHYP}. We can then identify the parameters $(c,a)$ of the pseudospheres with the parameters $(l,M)$ of the BTZ metric.

Taking into account that parameters of the elliptic pseudosphere ELL satisfy $0<c=a\sin\theta<a$, we obtain the correspondence
\begin{align}
    a &= l \,, \label{BTZ-to-pseudosphere-1}\\
    \text{ELL}: \qquad c &= l\,\sqrt{-M}=a\,\sqrt{-M}\, , &  -1&<M<0 \,, \label{BTZ-to-pseudosphere-2}\\
    \text{BEL}: \qquad c&= \beta, &  M&=0\,, \label{BTZ-to-pseudosphere-3}\\
    \text{HYP}: \qquad c &= l\,\sqrt{M}=a\,\sqrt{M}\,, & M&>0 \,. \label{BTZ-to-pseudosphere-4}
\end{align}

The pseudospherical surface is defined only where the embedding coordinate $z(u)\in \mathbb{R}$. Considering that we chose the parametrization, in which \eqref{arc-length-parametrization} is satisfied and $R(u)=\mathcal{R}(u)$, we obtain ranges of the $u$-coordinate for the profile functions \eqref{BTZ-profile-ELL}--\eqref{BTZ-profile-HYP} as
\begin{align}
    u & \in [0, l \arccosh \frac{1}{\sqrt{-M}} ] \,, \label{BTZ-u-range-ELL} \\
    u & \in (-\infty, l\,\ln{\frac{l}{\beta}}] \,, \label{BTZ-u-range-BEL} \\
    u & \in [-l \arccosh \sqrt{1+\frac{1}{M}} \,, l \arccosh \sqrt{1+\frac{1}{M}} ]  \,. \label{BTZ-u-range-HYP}
\end{align}
Using the dictionary \eqref{BTZ-to-pseudosphere-1}--\eqref{BTZ-to-pseudosphere-4} between the BTZ parameters and those of the pseudospheres, the ranges \eqref{BTZ-u-range-ELL}--\eqref{BTZ-u-range-HYP} coincide with the corresponding embedding domains for the three pseudospheres BEL, ELL and HYP given in \eqref{ELL-u-range}--\eqref{HYP-u-range}.

Therefore, depending on the mass parameter $M$, the spatial part of the BTZ spacetime is realized by the Euclidean embeddings of the three classical pseudospheres in $\mathbb{R}^3$~\cite{eisenhart1909treatise}, \cite{ovchinnikov}.

The main interest in considering ELL, BEL and HYP, is not purely theoretical; rather it is motivated by a specific realization of these geometries constructed from graphene, described in the laboratory by the metric
\begin{equation}\label{graphene-lab-metric}
    ds^2_{graph}=dt^2-du^2-R^2(u)\,dv^2 \,,
\end{equation}
where $R(u)$ is one of the profile functions \eqref{ELL}--\eqref{HYP}.

\begin{figure}[t]
    \centering
    \includegraphics[width=0.4\linewidth]{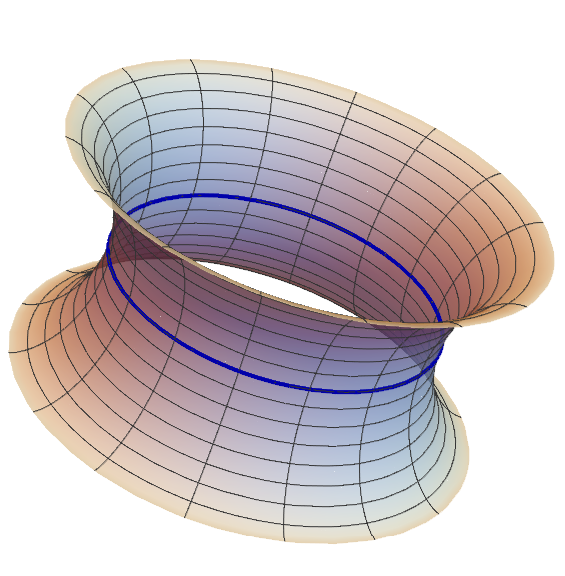}
\caption{The minimum radial distance $R(u)=c$ from the z-axis, corresponding to the waist of the HYP.}
    \label{fig: HYP waist}
\end{figure}

Although the coordinate $\mathcal{R}_H$ of the BTZ BH horizon \eqref{defRhor} coincides with the waist of the hyperbolic pseudosphere HYP, located at $R(u)=c$ depicted in Fig.~\ref{fig: HYP waist} (see  Eqs.~\eqref{defRhor}, \eqref{identification}, \eqref{BTZ-to-pseudosphere-1} and \eqref{BTZ-to-pseudosphere-4} to relate the parameters), this correspondence is purely geometrical. Indeed, starting with the full static, non-rotating BTZ metric \eqref{BTZ-metric} and applying the coordinate transformation \eqref{du-dR-BTZ}, we obtain
\begin{equation}
    ds^2_{\text{BTZ}}=\left(\frac{\mathcal{R}^2(u)}{l^2}-M\right) \, d\mathcal{T}^2-du^2-\mathcal{R}^2(u)\,d\mathcal{V}^2 \,.
\end{equation}
For the static BTZ solution, the event horizon coincides with the KH and is located where the norm of the timelike Killing vector, $\xi=\partial_{\mathcal{T}}$, vanishes,
\begin{equation}
    g_{\mathcal{T}\mathcal{T}}=\frac{\mathcal{R}^2}{l^2}-M=0 \,.
\end{equation}
We obtain the horizon radius $\mathcal{R}_H=l\,\sqrt{M}$. By contrast, the graphene laboratory metric \eqref{graphene-lab-metric} satisfies
\begin{equation}
    g_{tt}=1 \ \ \text{everywhere,}
\end{equation}
and therefore does not possess a KH.

More generally, one can consider broader notions of horizons in relativistic physics and examine the existence of trapped surfaces. For the ultrastatic metric \eqref{graphene-lab-metric}, with outgoing and ingoing null directions generated by $k_{\pm}=\partial_t\pm\partial_u$, the corresponding null expansions are easily calculated to be
\begin{equation}
    \theta_{\pm}=\pm\frac{R'(u)}{R(u)}\,.
\end{equation}

For $R'(u)\neq 0$, the expansions $\theta_+$ and $\theta_-$ have opposite signs, so one future-directed null congruence is expanding while the other is contracting. Hence, no trapped surfaces occur, and no apparent horizon forms.

The special case occurs for the profile function \mbox{$R(u)=c\cosh (u/a)$} of the hyperbolic pseudosphere HYP. Since $R'(0)=0$, both null expansions vanish at $u=0$:
\begin{equation}
    \theta_{\pm}=0 \,.
\end{equation}
However, there is still no trapping horizon, since there is no region in which both future-directed null expansions are non-vanishing and have the same sign. The HYP profile satisfies
\begin{equation}
    R''(0)=\frac{c}{a^2}>0 \,,
\end{equation}
showing that the circle $u=0$ is a local minimum of the function $R(u)$, and therefore corresponds to a throat of the spatial geometry.

This observation motivates us to explore a wormhole analogy, since in general relativity traversable wormholes are also characterized by vanishing null expansions at the throat.

\section{Realization of a wormhole spacetime metric using a hyperbolic pseudosphere}\label{Wormhole-section}

Motivated by the structural analogy between the geometry of the hyperbolic pseudosphere and that of a wormhole spatial bridge (a central throat connecting two asymptotic regions), we establish a mapping between the full (2+1)-dimensional Lorentzian spacetime and the pseudospherical metric coordinates.

We consider a standard Morris-Thorne wormhole line element \cite{morris1988} restricted to the equatorial plane ($\theta=\frac{\pi}{2}$) and assuming a vanishing redshift function, $\Phi(r)=0$. The resulting static spacetime metric reads:
\begin{equation}\label{wormhole-metric}
    ds^2_{wh}=dt^2-\frac{dr^2}{1-\frac{b(r)}{r}}-r^2\,d\phi^2 \ ,
\end{equation}
where $b(r)$ represents the shape function defining the spatial embedding profile.

The equivalent line element generated on a carbon-made surface of revolution possessing the geometry of a hyperbolic pseudosphere (HYP) is described by:
\begin{equation}\label{HYP-metric}
    ds^2_{graph}=dt^2-du^2-R^2(u)\,dv^2, \quad R(u)=c\cosh\left(\frac{u}{a}\right) .
\end{equation}
The line elements \eqref{wormhole-metric} and \eqref{HYP-metric} map identically onto each other under the coordinate identifications:
\begin{equation}\label{wormhole-id}
    v=\phi,  \quad du=\frac{dr}{\sqrt{1-b(r)/r}}, \quad R(u)=r(u) \ .
\end{equation}
Given the profile function $R(u)=c\cosh(u/a)$ of the HYP surface, the corresponding shape function $b(r)$ is determined by the differential relation in \eqref{wormhole-id}. Differentiating $R(u)=r(u)$ yields $\frac{dr}{du}=\frac{c}{a}\sinh(u/a)$, which uniquely determines the effective wormhole shape function:
\begin{equation}\label{wormhole-shape}
    b(r)=r \left[1+\frac{c^2}{a^2}-\frac{r^2}{a^2}\right] \ .
\end{equation}

The throat of a traversable wormhole is located at a radial coordinate $r_0$ defined by the boundary condition $b(r_0)=r_0$ \cite{morris1988}. Applying this constraint to the shape function \eqref{wormhole-shape} leads to:
\begin{equation}
    b(r_0)=r_0\left[ 1+\frac{c^2}{a^2} -\frac{r_0^2}{a^2}\right] =r_0 \ ,
\end{equation}
which is identically satisfied when $r_0=c$. This choice isolates the minimal waist of the pseudosphere, $R(0)=c$, as the physical location of the spatial throat.

We next verify the geometric flare-out condition, which requires $b'(r_0)<1$ at the throat radius \cite{morris1988,Visser1995}. Differentiating Eq.~\eqref{wormhole-shape} and evaluating it at $r_0=c$ yields:
\begin{equation}
    b'(r_0=c)=1-2\frac{c^2}{a^2}<1 \ .
\end{equation}
The shape function \eqref{wormhole-shape} satisfies the flare-out criterion, confirming that the coordinate $r_0=c$ represents a strict minimum of the spatial radius and that the coordinate geometry opens outward symmetrically on both sides of the throat. Because this carbon-based framework operates as an analog gravity model, the enforcement of the flare-out condition does not imply constraints on physical energy conditions; its validity remains entirely geometric.

Consequently, the identification yields a full spacetime equivalence between the two metrics:
\begin{equation}
    ds^2_{graph}\equiv ds^2_{wh} \ , \quad \text{for} \quad \Phi(r)=0 \,.
\end{equation}
This match represents a structural line-element equivalence across the entire coordinates, rather than a restricted spatial slice or a Weyl-conformal correspondence.

The effective spacetime metric $ds^2_{graph}$ is engineered directly via the structural arrangement of the carbon membrane and does not emerge as a dynamical solution of Einstein field equations. While the graphene architecture reproduces the kinematics of an ultrastatic traversable wormhole background, it does not incorporate gravitational dynamics or backreaction phenomena. The carbon membrane operates as a fixed geometric background that does not deform in response to the motion of the Dirac quasiparticles. The wormhole shape function \eqref{wormhole-shape} must therefore be interpreted as an effective geometry that is decoupled from an analogue stress-energy tensor. This configuration realizes an ultrastatic traversable wormhole framework where the kinematics of spinorial light rays can be systematically analyzed by tracking the exact geodesics of massless fermions.

\section{Semiclassical Strategy: ray-tracing and 3d Raychaudhuri equation}
\label{sec:strategy}

Before proceeding with the evaluation of the traversability of the carbon BTZ wormhole, we establish the framework to map the multi-channel transport problem onto the behavior of uncoupled classical geodesics. This formulation addresses the elimination of both the carrier's spin and the collective distortions of the beam, reducing the transport analysis to a ray-tracing description for spinless, massless particles.

The decoupling of the spinor degrees of freedom from the trajectories is achieved by invoking a short-wavelength eikonal approximation ($\lambda_e \ll a,c$). This is derived in Appendix D, via a perturbative WKB expansion \cite{rubinow1963}, adapted to our emergent set-up. As discussed in Appendix D, we choose the parameter of the expansion to be $\epsilon=\lambda_e/R_{macro}$, with $R_{macro} = a, c$.

The leading-order $\mathcal{O}(1/\epsilon)$ algebraic consistency of the Dirac equation requires the matrix operator to be non-invertible to sustain a physical current. This condition yields a scalar, spin-independent Hamilton-Jacobi equation
\begin{equation}
    g^{\mu\nu}(x) \partial_\mu S \partial_\nu S = 0 \,.
\end{equation}
Furthermore, within the framework of a torsion-free geometry, any potential non-geodesic driving terms are naturally constrained. At the leading eikonal order, the spinorial and torsional degrees of freedom decouple from the path determination, ensuring that the charge carriers strictly track the pure metric geodesics of the hyperbolic manifold, as derived via the WKB expansion in Appendix D.

Within these limits, individual excitations are described as classical rays. Let us now determine whether a beam of such classical particles can be analyzed via independent trajectories rather than an interacting geometric congruence. To do so, we implement the Raychaudhuri equation, which governs the kinematic evolution of a geodesic bundle characterized by the expansion scalar $\theta$, the shear tensor $\sigma_{\mu\nu}$, and the vorticity tensor $\omega_{\mu\nu}$ \cite{witten2020}. For a null geodesic congruence with affine parameter $\lambda$ and tangent vector field $k^\mu = dx^\mu / d\lambda$, the general Raychaudhuri equation reads \cite{kar2007}
\begin{equation} \label{eq:raychaudhuri_general}
\frac{d\theta}{d\lambda} = -\frac{1}{d-2}\theta^2 - \sigma_{\mu\nu}\sigma^{\mu\nu} + \omega_{\mu\nu}\omega^{\mu\nu} - R_{\mu\nu}k^\mu k^\nu \,,  \end{equation}
where $d$ is the spacetime dimension.

In a $d=3$ framework, the subspace transverse to the direction of propagation is one-dimensional. Because a one-dimensional vector space does not accommodate asymmetric deformations or rotations, the transverse shear and vorticity tensors vanish structurally ($\sigma_{\mu\nu} \equiv 0$ and $\omega_{\mu\nu} \equiv 0$) for any null congruence \cite{kar2007,witten2020}. This constitutes a geometric property of low-dimensional spacetimes. Consequently, Eq.~\eqref{eq:raychaudhuri_general} reduces to:
\begin{equation}
    \frac{d\theta}{d\lambda} = -\theta^2 - R_{\mu\nu}k^\mu k^\nu \,.
    \label{eq:raychaudhuri_reduced}
\end{equation}

While the general structure of Eq.~\eqref{eq:raychaudhuri_reduced} is determined by the low-dimensional spacetime geometry, the geometric focusing or defocusing of the charge carriers depends on the specific curvature profile of the surface. To evaluate this explicitly, we use the metric \eqref{HYP-metric}. Due to its ultrastatic nature, the non-vanishing components of the Riemann tensor, $R^\alpha_{\;\beta\gamma\delta}$, and the Ricci tensor, $R_{\mu\nu}$, originate strictly from the two-dimensional spatial metric of the carbon sheet. A direct calculation of the Ricci tensor components in this signature yields
\begin{equation}
    R_{tt} = 0, \quad R_{uu} = -\frac{R''(u)}{R(u)} = -\frac{1}{a^2}, \quad R_{vv} = -R(u) R''(u) = -\frac{c^2}{a^2}\cosh^2\left(\frac{u}{a}\right) \,.
\end{equation}

We now project this tensor structure onto the null geodesic vector field $k^\mu = (\omega, k^u, k^v)$, where $\omega = dt/d\lambda$ represents the conserved excitation energy. The null condition $g_{\mu\nu}k^\mu k^\nu = 0$ implies $\omega^2 - (k^u)^2 - R^2(u)(k^v)^2 = 0$, which yields $\omega^2 = (k^u)^2 + R^2(u)(k^v)^2$. Evaluating the geometric driving term $R_{\mu\nu}k^\mu k^\nu$ results in
\begin{equation}
    R_{\mu\nu}k^\mu k^\nu = R_{uu} (k^u)^2 + R_{vv} (k^v)^2 = -\frac{1}{a^2} (k^u)^2 - \frac{R^2(u)}{a^2} (k^v)^2 = -\frac{1}{a^2} \left[ (k^u)^2 + R^2(u)(k^v)^2 \right] \,.
\end{equation}
By virtue of the null constraint, the bracketed spatial term is identically equal to $\omega^2$. Therefore, the contraction simplifies to a negative contribution proportional to the square of the energy:
\begin{equation}
    R_{\mu\nu}k^\mu k^\nu = -\frac{\omega^2}{a^2} \,.
\end{equation}

Substituting this expression into the $(2+1)$-dimensional reduced Raychaudhuri equation \eqref{eq:raychaudhuri_reduced}, we obtain the exact kinematic evolution equation for the bundle expansion:
\begin{equation}
    \frac{d\theta}{d\lambda} = -\theta^2 + \frac{\omega^2}{a^2} \,.
    \label{eq:ray_final_riccati}
\end{equation}
Equation \eqref{eq:ray_final_riccati} is a Riccati differential equation, where the curvature term enters with a positive sign, $+ \omega^2 / a^2$, due to the intrinsic negative geometry of the HYP background. In standard general relativity, an attractive gravitational field introduces a negative driving term via the strong energy condition, forcing $\frac{d\theta}{d\lambda} \le -\theta^2$, which inevitably triggers a singular collapse ($\theta \to -\infty$) and creates a caustic. Here, the opposite occurs: the intrinsic negative curvature acts instead as a defocusing geometric mechanism.

To formally connect this behavior to the single-geodesic framework, we perform the transformation $\theta = \frac{1}{\eta}\frac{d\eta}{d\lambda}$, where $\eta(\lambda)$ represents the transverse separation (the Jacobi field) between neighboring rays within the bundle. Under this substitution, the non-linear Riccati equation \eqref{eq:ray_final_riccati} linearizes perfectly into a regular Jacobi deviation equation:
\begin{equation}
    \frac{d^2\eta}{d\lambda^2} - \frac{\omega^2}{a^2}\eta = 0 \,.
    \label{eq:ray_jacobi_final}
\end{equation}
The general solution to Eq.~\eqref{eq:ray_jacobi_final} is a linear combination of hyperbolic functions, $\eta(\lambda) = A \cosh(\omega \lambda / a) + B \sinh(\omega \lambda / a)$. Since $\theta(0)=(\omega/a)\,B/A$, the Jacobi field $\eta(\lambda)$ can vanish for $\lambda>0$ only if $B/A<-1$, i.e.\ only if the bundle is initially converging faster than $\theta(0)=-\omega/a$. For any bundle with $\theta(0)>-\omega/a$ (in particular for initially parallel or diverging beams), $\eta(\lambda)$ never undergoes zero-crossings inside the hyperbolic domain, the expansion $\theta$ remains finite, and no caustics (focal points) or intersecting ray envelopes are generated by the curvature of the HYP: the negative curvature can only delay, never induce, focusing.

This geometric regularity indicates that individual geodesic paths remain ordered and non-intersecting throughout the spatial throat, justifying the evaluation of the total classical traversability by integrating over single, isolated geodesic rays.

\section{Geodesics for light-like quasiparticles}

In this section, we evaluate the geodesics of the Dirac quasiparticles within the analogue system described by the ultrastatic line element:
\begin{equation}\label{ultrastatic-graphene-metric}
    ds_{graph}^2=dt^2-du^2-R^2(u)\,dv^2 \equiv dt^2-h_{ij}\,dx^i\,dx^j \,,
\end{equation}
where $h_{ij} = \text{diag}(1, R^2(u))$ defines the positive-definite Riemannian metric of the spatial carbon membrane. Near the charge neutrality point, the charge carriers are treated as light-like excitations propagating at the effective Fermi velocity, which is set to $v_F=1$.

The trajectories are determined by solving the geodesic equations associated with the metric \eqref{ultrastatic-graphene-metric}. Due to the ultrastatic framework, the only non-vanishing Christoffel symbols originate strictly from the spatial components $h_{ij}$:
\begin{align}
    &\Gamma^t_{\mu\nu}=0 \,, \qquad \Gamma^i_{tt}=0 \,, \\
    &\Gamma^u_{vv}=-R(u)R'(u) \,, \qquad \Gamma^v_{uv}=\Gamma^v_{vu}=\frac{R'(u)}{R(u)} \label{christoffels} \ ,
\end{align}
where $R'(u)\equiv\frac{dR}{du}$. The geodesic differential relation for an affinely parametrized curve $x^{\mu}(\lambda)=\left(t(\lambda), x^i(\lambda)\right)$ splits into independent temporal and spatial components:
\begin{align}
    &\frac{d^2t}{d\lambda^2}=0 \,, \label{temporal-geodesic-eq}\\
    &\frac{d^2x^i}{d\lambda^2}+\Gamma^i_{jk}\frac{dx^j}{d\lambda}\frac{dx^k}{d\lambda}=0 \label{spatial-geodesic-eqn} \,.
\end{align}
Equation \eqref{temporal-geodesic-eq} indicates that the coordinate time $t$ serves as an affine parameter along the trajectories. Consequently, the curve can be parameterized as $x^\mu(t)$ via the affine transformation $t=\alpha\lambda+t_0$, with $\alpha, t_0$ constants. Furthermore, because the quasiparticles are light-like, the condition $ds^2_{graph} = 0$ in Eq.~\eqref{ultrastatic-graphene-metric} implies:
\begin{equation}
    dt^2=h_{ij}\,dx^i\,dx^j\equiv dl^2 \,,
    \label{eq:arc_length_null}
\end{equation}
where $dl$ represents the spatial arc length. Choosing the positive branch for the parameter mapping, Eq.~\eqref{eq:arc_length_null} implies $\frac{d^2l}{d\lambda^2}=0$, confirming that $l$ operates as an affine parameter. The spatial trajectory equation \eqref{spatial-geodesic-eqn} can thus be rewritten as:
\begin{equation}\label{geodesic-eqn-final}
     \frac{d^2 x^i}{dl^2}+\Gamma^i_{jk} \frac{dx^j}{dl}\frac{dx^k}{dl}=0 \,,
\end{equation}
which corresponds to the geodesic equation on the static Riemannian spatial slice.

The surface of revolution representing the carbon membrane is described by the metric:
\begin{equation}
      dl^2 = h_{ij} dx^i dx^j=du^2+R^2(u)dv^2 \,,
\end{equation}
where $R(u)$ is the profile function of the pseudosphere. This geometry allows the analysis of the light-like trajectories to be conducted within the intrinsic framework of two-dimensional surfaces of revolution.

Substituting the Christoffel symbols \eqref{christoffels} into the reduced geodesic equation \eqref{geodesic-eqn-final} yields the coupled system:
\begin{align}
    &\ddot{v}+2 \ \frac{R'(u)}{R(u)}\dot{u}\dot{v}=0 \label{geodesic-eqn-1} \,, \\
    &\ddot{u}-R(u)R'(u) \ \dot{v}^2=0 \label{geodesic-eqn-2} \,,
\end{align}
where the dots denote differentiation with respect to the spatial arc length parameter, $\dot{u}\equiv\frac{du}{dl}$ and $\dot{v}\equiv\frac{dv}{dl}$.

Integrating Eq.~\eqref{geodesic-eqn-1} yields a first integral of the motion:
\begin{equation}\label{integral-1}
    R^2(u)\ \dot{v}=J=\text{const}\ ,
\end{equation}
where $J$ is the conserved angular momentum, representing the Clairaut constant of the surface. According to Clairaut's theorem \cite{doCarmo1976}, this relation satisfies:
\begin{equation}\label{Clairaut}
    |J|=R\sin\theta\,,
\end{equation}
where $\theta$ is the angle between the tangent vector to the geodesic and the local meridian of the surface of revolution.

Combining the first integral \eqref{integral-1} with the unit tangent vector normalization condition, $\dot{u}^2+R^2(u)\,\dot{v}^2=1$, leads to the radial constraint:
\begin{equation}\label{integral-2}
    \dot{u}^2=1-\frac{J^2}{R^2(u)} \ .
\end{equation}
Equation \eqref{integral-2} represents the effective energy conservation equation for the radial coordinate. Differentiating Eq.~\eqref{integral-2} reproduces the second spatial geodesic relation \eqref{geodesic-eqn-2}, confirming that the system is fully determined by the first integrals. Applying the chain rule $\dot{u}=\frac{du}{dv}\dot{v}$ in combination with Eqs.~\eqref{integral-1} and \eqref{integral-2} provides the final differential equation for the geodesic paths:
\begin{equation}\label{geodesics-integral}
     \frac{dv}{du}=\pm \frac{J}{R(u)\sqrt{R^2(u)-J^2}} \,.
\end{equation}

\subsection{Classification of the geodesics for the HYP}\label{geodesics-HYP-classification-section}

The motion on the surface of revolution in the meridional direction is determined by $\dot{u}$. Whenever $\dot{u}=0$, the geodesic reaches a turning point and reverses its direction of motion along the meridian. Imposing the turning-point condition, $\dot{u}=0$, in \eqref{integral-2}, we obtain
\begin{equation}
    \dot{u}^2=1-\frac{J^2}{R^2(u)}=0 \,.
\end{equation}
Solving this equation yields the condition for the turning point~$u_*$
\begin{equation}\label{turning-point-condition}
    R(u_*)=|J| \,.
\end{equation}

As shown in \secref{Wormhole-section}, the hyperbolic pseudosphere HYP with the profile function
\begin{equation}
    R(u)=c\cosh\ua
\end{equation}
is central to the investigation of the throat geometry.

The geodesics on the HYP are classified by the values of their angular momentum $|J|$. Depending on whether the turning-point condition \eqref{turning-point-condition} can be satisfied over the range of the profile function $R(u)$, they are divided into two classes.

The profile function is continuous and takes values in the interval
\begin{equation}
    R(u)\in [c,\sqrt{c^2+a^2}] \,.
\end{equation}
The classification of the geodesics is then completely determined by the value of $|J|$ relative to the minimum radius of the geometry $R_{min}=c$, i.e. the throat.

If $|J|<c$, the turning-point condition \eqref{turning-point-condition} cannot be satisfied by the geodesic, because $|J|$ lies below the minimum value, $R_{min}$, of the profile function. Therefore, the geodesic has no turning point, crosses the throat at $R_{min}=c$, and emerges on the opposite side of the HYP. We refer to such geodesics as \
\textit{traversing} geodesics.

If $|J| > c$, the value of $|J|$ belongs to the range of the profile function, and therefore the turning-point condition \eqref{turning-point-condition} is satisfied for some $u_*$. In this case, the geodesic reaches the turning point at $R(u_*)$, reverses its direction of motion, and returns to the same side of the HYP. Since the geodesic returns to the same side of the HYP, we refer to such geodesics as \mbox{\textit{reflecting}} geodesics. The limiting case $|J|=c$ (critical geodesics) is special: since $\dot u^2 = 1 - c^2/R^2(u) \simeq u^2/a^2$ near $u=0$, the throat is approached only asymptotically, after an infinite arc length, so that critical geodesics neither cross the throat nor turn back at finite length (see Sec.~8).

\begin{figure}[t]
    \centering
    \includegraphics[width=0.3\linewidth]{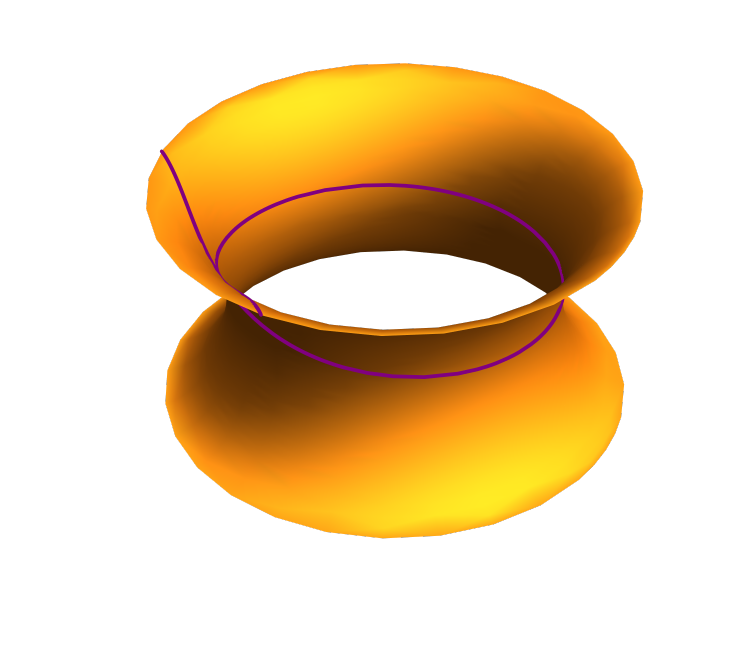}
    \includegraphics[width=0.3\linewidth]{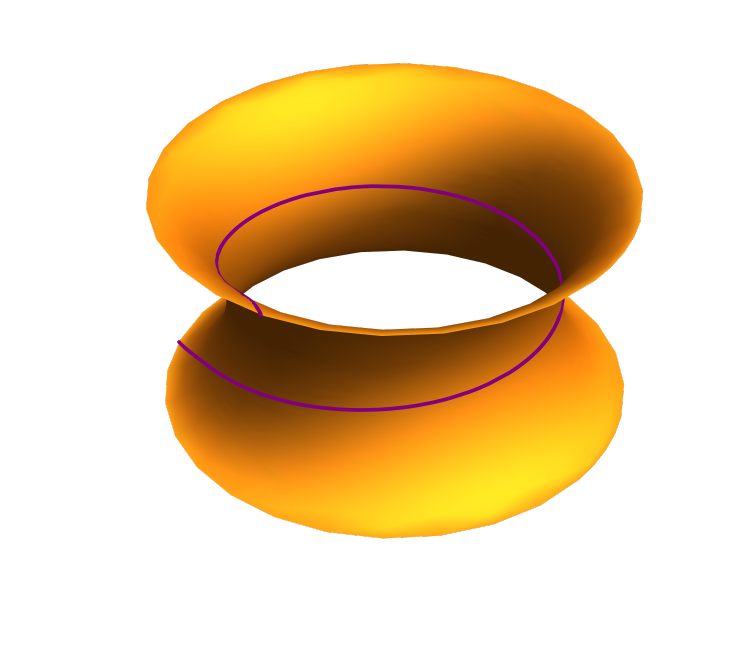}
    \caption{Geodesics on the HYP for $c=1$ and $|J|=1.0005$ (left) and $|J|=0.9995$ (right).}
    \label{HYP-geodesics-c-1}
\end{figure}

\begin{figure}[t]
    \centering
    \includegraphics[width=0.3\linewidth]{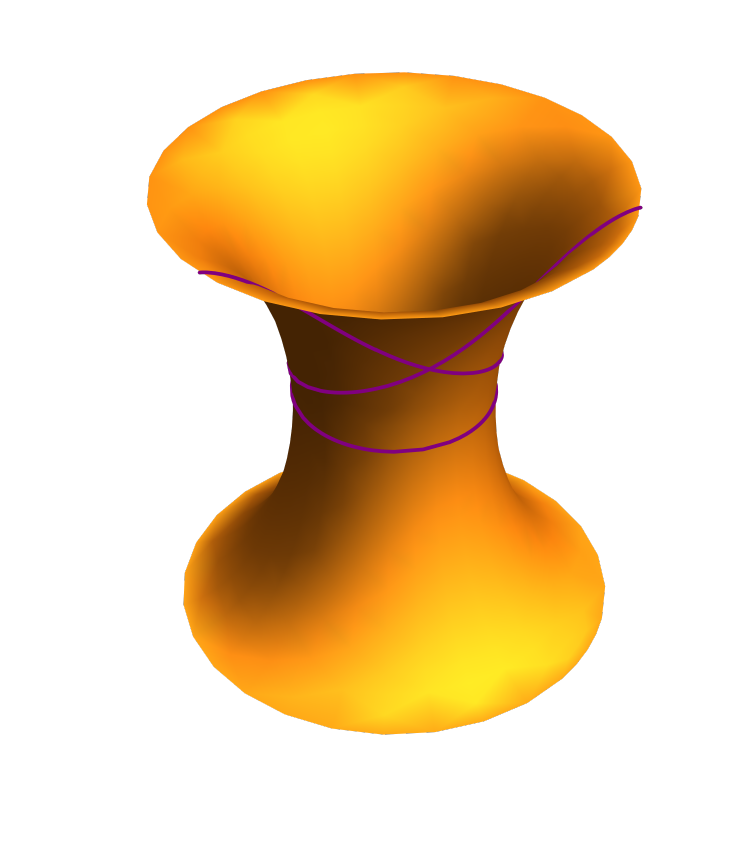}
    \includegraphics[width=0.3\linewidth]{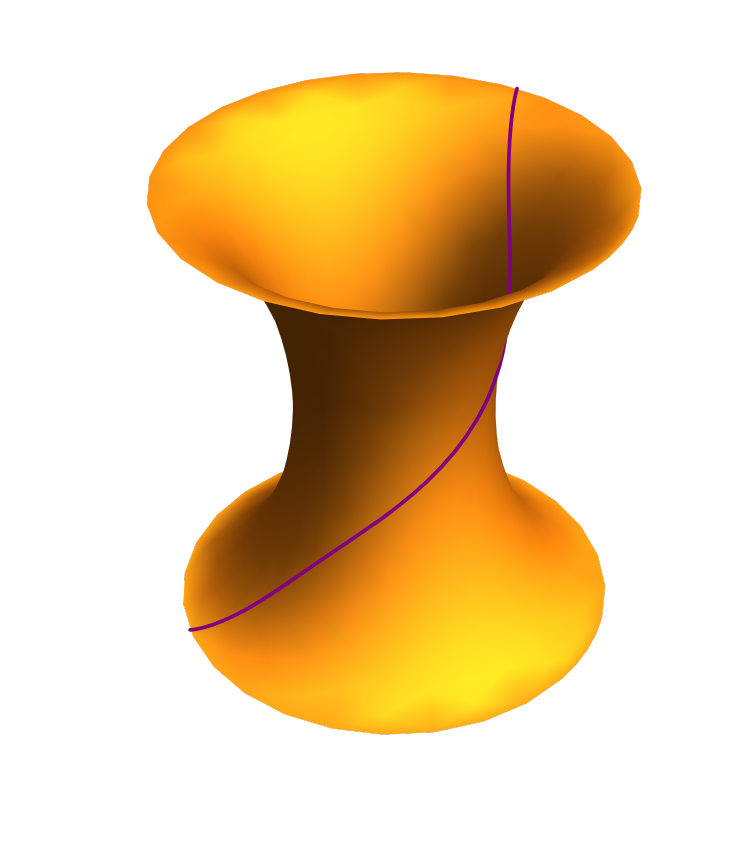}
    \caption{Geodesics on the HYP for $c=0.5$ and $|J|=0.5005$ (left) and $|J|=0.3870$ (right).}
    \label{HYP-geodesics-c-05}
\end{figure}

\begin{figure}
    \centering
    \includegraphics[width=0.3\linewidth]{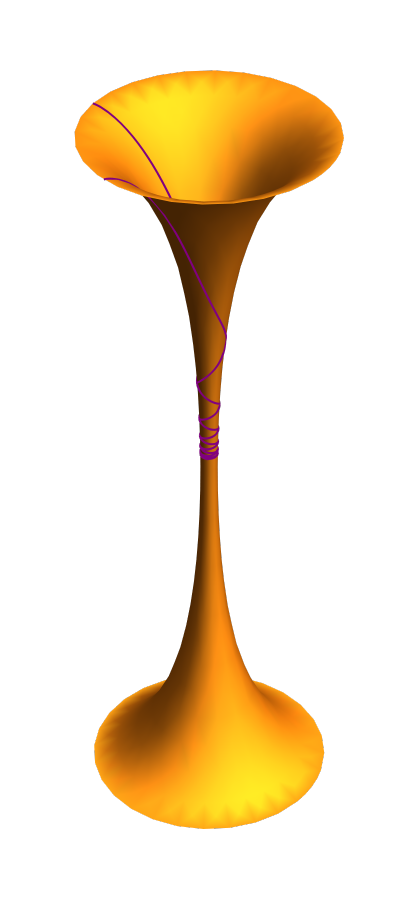}
    \includegraphics[width=0.3\linewidth]{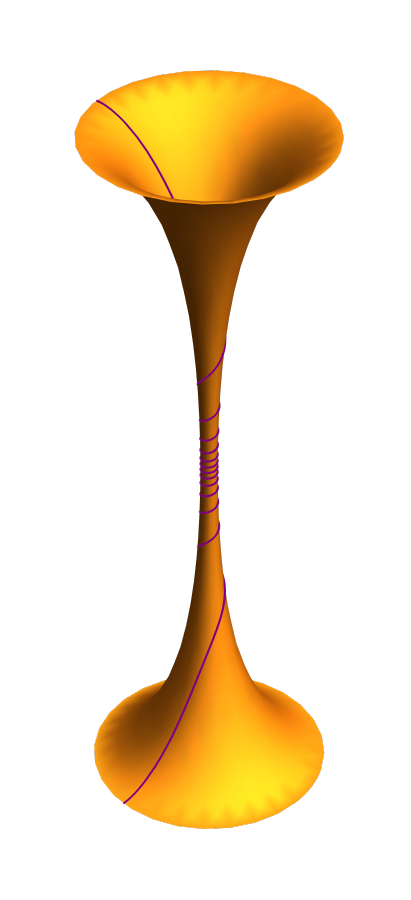}
    \caption{Geodesics on the HYP for $c=0.07$ and $|J|=0.0705$ (left) and $|J|=0.0695$ (right).}
    \label{HYP-geodesics-c-007}
\end{figure}

\subsection{Geodesics for the HYP}\label{geodesics-HYP-section}

To find the geodesic path on the surface, we integrate \eqref{geodesics-integral}
\begin{equation}\label{geodesics-integral-2}
    v(u)=\pm\int\frac{J\,du}{R(u)\sqrt{R^2(u)-J^2}} +C\,,
\end{equation}
where the integration constant $C$ is determined by the initial conditions $(v_0, u_0)$ for the geodesic.

Choosing the positive orientation of the geodesics and substituting the profile function $R(u)=c\cosh\ua$ for the HYP, we obtain the indefinite integral
\begin{equation}\label{geodesics-indefite-integral-HYP}
    I(u)=\int\frac{J\,du}{c\cosh\ua\sqrt{c^2\cosh^2\ua-J^2}} \,.
\end{equation}
We introduce the substitution $t=\tanh\ua$ to rewrite the integral in terms of the algebraic functions,
\begin{equation}
    I(u)=\frac{a}{c}\int\frac{J\,dt}{\sqrt{c^2-J^2+J^2 t^2}}\,.
\end{equation}

Since there are two types of geodesics (traversing and reflecting), classified by the absolute value of the angular momentum, $|J|$, we consider two cases for the denominator
\begin{align}
    &|J|<c: \qquad I(u)=\frac{a}{c}\int\frac{J\,dt}{\sqrt{A^2+J^2t^2}}, \quad A\equiv\sqrt{c^2-J^2} \,, \\
    &|J|>c: \qquad I(u)=\frac{a}{c}\int\frac{J\,dt}{\sqrt{J^2t^2-A^2}}, \quad A\equiv \sqrt{J^2-c^2}\,.
\end{align}

Therefore, by evaluating the above antiderivatives, we obtain the two families of geodesics (where we restrict our analysis to the positive branch $t = \tanh(u/a) > 0$, without loss of generality, due to the reflection symmetry of the pseudospherical profile around the throat $u=0$).

\begin{align}
    &|J|<c: \quad v_T(u)=\frac{a}{c}\arcsinh \left( \frac{J\tanh\ua}{\sqrt{c^2-J^2}}\right) + C \,, \\
    &|J|>c: \quad v_R(u)=\frac{a}{c}\sgn(J)\arccosh \left( \frac{|J||\tanh\ua|}{\sqrt{J^2-c^2}} \right) + C \,,
\end{align}
where subscripts $T$ and $R$ denote traversing and reflecting geodesics, respectively.

To illustrate both types of geodesics, we plot selected trajectories for various values of the pseudosphere parameter $c$ and the magnitude of the angular momentum $|J|$, as shown in Figs.~\ref{HYP-geodesics-c-1}--\ref{HYP-geodesics-c-007}. An interactive tool for exploring the geodesics is provided in the Supplemental Material, allowing readers to vary both parameters and visualize the corresponding geodesics on the HYP.

\section{Gravitational lensing setup}

\begin{figure}
    \centering
    \includegraphics[width=0.6\linewidth]{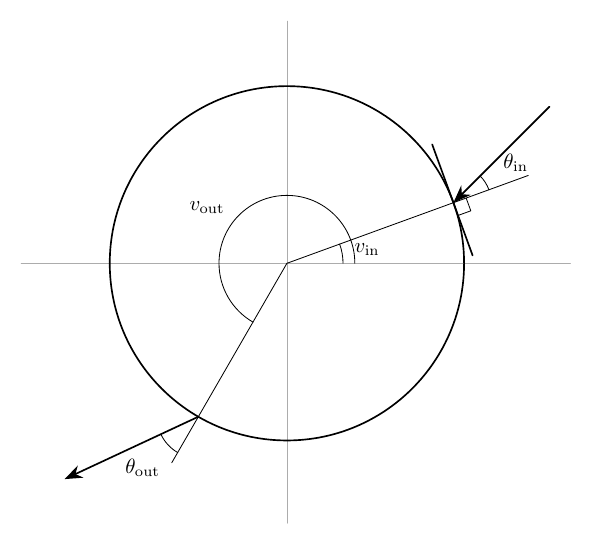}
    \caption{Geometry of the incoming and outgoing rays in the gravitational lensing setup.}
    \label{Lensing_setup_angles}
\end{figure}

The throat geometry of the HYP (with $R(u)=c\cosh\ua$) allows us to realize gravitational lensing in the graphene analogue system. To investigate this, we consider a graphene-based HYP geometry smoothly connecting two flat graphene sheets. The geodesic equations \eqref{geodesic-eqn-1} and \eqref{geodesic-eqn-2} allow us to calculate the deflection angle $\delta$ of a graphene quasiparticle in the ray approximation (see Appendix \ref{app:WKB} for details) as it propagates from one of the flat graphene sheets into the HYP geometry. Specifically, we consider rays originating from the upper sheet and propagating downward the HYP throat.

As illustrated in Fig.~\ref{Lensing_setup_angles}, in the top-down view along the z-axis, the incoming ray approaches the outer circular boundary of the pseudosphere, located at a radial distance $R_\mathrm{max}$ from the z-axis, at an incidence angle $\theta_\mathrm{in}$ measured from the local normal to the boundary. The ray enters the HYP geometry at the angular coordinate $v_\mathrm{in}$ and follows either a traversing or a reflecting geodesic. In the former case, the ray passes through the throat and emerges into the lower flat graphene sheet, whereas in the latter, it returns to the upper sheet. In both cases, the ray exits the HYP at the angular coordinate $v_\mathrm{out}$, making an exit angle $\theta_\mathrm{out}$ with the local normal to the boundary circle of radius $R_\mathrm{max}$.

The total deflection angle $\delta$ between the incoming and outgoing asymptotic rays in the flat graphene sheets, taking orientations into account, can then be expressed as
\begin{equation}\label{deflection-angle}
    \delta=|v_{\mathrm{out}}-v_{\mathrm{in}}|-(\theta_{\mathrm{in}}-\theta_{\mathrm{out}}) \,.
\end{equation}

Clairaut theorem, introduced above in connection with the conserved angular momentum $J$ of the geodesic Eqn.~\eqref{integral-1}, can be used to determine the incidence and exit angles $\theta_{\mathrm{in}}$ and $\theta_{\mathrm{out}}$ in the graphene sheets. As a first integral of the geodesic equations, Clairaut's theorem holds wherever the surface is smooth and differentiable. In particular, near the rim at $R_{\mathrm{max}}$, the hyperbolic membrane smoothly connects to the flat graphene sheets, with its tangent plane aligning with that of the adjoining planar geometry. Applying Clairaut's theorem \eqref{Clairaut} at the outer boundary of the HYP, we obtain
\begin{equation}
    \sin \theta_{\text{in,out}} = \frac{|J|}{R_{\text{max}}},
    \label{Clairaut-at-rim}
\end{equation}
where $\theta_{\text{in}}$ and $\theta_{\text{out}}$ represent the scattering angles measured with respect to the normal of the boundary circle.

For our gravitational lensing setup, we assume that the ray is incident on the upper plane and falls into the pseudosphere, advancing in the direction of the decreasing $u$-coordinate.

The sign of $\dot{u}$ determines whether the geodesic moves toward increasing ($\dot{u}>0$) or decreasing ($\dot{u}<0$) values of $u$. In \secref{geodesics-HYP-classification-section}, we found that the HYP possesses two types of geodesics -- traversing and reflecting. Since traversing geodesics do not possess a turning point, the sign of $\dot{u}$ remains unchanged along their entire path. On the other hand, reflecting geodesics change the sign of $\dot{u}$ after reaching the turning point.

Therefore, traversing geodesics satisfy $\theta_{\mathrm{in}}=\theta_{\mathrm{out}}$. For reflecting geodesics, Eq.~\eqref{Clairaut-at-rim} gives
\begin{equation}\label{theta-in-out-result}
    \theta_{\mathrm{in}}=\pi-\arcsin\frac{|J|}{R_{\max}}\,, \qquad \theta_{\mathrm{out}}=\arcsin\frac{|J|}{R_{\max}}.
\end{equation}

\subsection{Total azimuthal sweep $|\Delta v_{\mathrm{tot}}|$}
To calculate the deflection angle \eqref{deflection-angle}, we need the angular sweep
\begin{equation}
    |\Delta v_{\mathrm{tot}}|=|v_{\mathrm{out}}-v_{\mathrm{in}}|,
\end{equation}
of the geodesic along the pseudosphere, between the entrance and exit points on the rim $R=R_{\max}=\sqrt{a^2+c^2}$, located at $u=\pm u_{\max}$, $u_{\max}=a\,\arccosh\sqrt{1+a^2/c^2}$ (see Eq.~\eqref{HYP-u-range}).

Because of symmetry, each geodesic consists of two identical segments on either side of its point of closest approach, $u_\star$ (which is either the throat, $u_\star=0$, or the turning point, $u_\star=u_*$). Therefore, it is sufficient to compute only half of the angular sweep and multiply the result by two to obtain the total angular sweep along the surface:
\begin{equation}
    |\Delta v_{\mathrm{tot}}|=2 \,|I| = 2 \,\int_{u_{\star}}^{u_{\max}}\frac{|J| d\bar{u}}{R(\bar{u})\sqrt{R^2(\bar{u})-J^2}} \,.
\end{equation}

To evaluate the definite integral for the angular half-sweep, $|I|$, we rewrite it in terms of the profile function $R(u)$, obtaining
\begin{equation}\label{half-sweep-HYP}
    |I|= \int_{R(u_{\star})}^{R_{\max}} \frac{|J|\,a \,dR}{R\sqrt{R^2-J^2}\,\sqrt{R^2-c^2}} \,.
\end{equation}

Since the HYP possesses two types of geodesics, we need to calculate the half-sweep integral \eqref{half-sweep-HYP} separately for each type.

\subsubsection{Half-sweep for traversing geodesics ($|J|<c$)}
For traversing geodesics, the point of closest approach is the throat of the HYP, $u_\star=0$, corresponding to $R(u_\star)=c$. Therefore, the half-sweep integral is
\begin{equation}\label{half-sweep-crossing}
    |I|_T=\int_c^{R_{\max}} \frac{|J|\,a\, dR}{R\sqrt{R^2-J^2}\sqrt{R^2-c^2}} \,.
\end{equation}

The integrand in \eqref{half-sweep-crossing} becomes singular at the lower limit $R=c$, but the integral is still finite. To remove this singularity, we introduce the substitution $R=c\cosh y$, which gives
\begin{equation}
    |I|_T=\frac{a}{c}\int_0^{y_T}\frac{|J|\,dy}{\cosh y \sqrt{c^2\cosh^2 y-J^2}} \,.
\end{equation}

where $\cosh y_T = R_{\max}/c$. Introducing the hyperbolic substitution $t=\tanh y$, with $t_T\equiv\tanh y_T = a/\sqrt{a^2+c^2}$, we obtain\footnotemark
\begin{equation}\label{half-sweep-crossing-result}
    |I|_T=\frac{a}{c}\int_0^{t_T}\frac{|J|\,dt}{\sqrt{c^2-J^2+J^2t^2}}=\frac{a}{c}\arctanh\left(\frac{a|J|}{c\sqrt{a^2+c^2-J^2}}\right) \,,
\end{equation}
\footnotetext{We used the identity $\arcsinh z=\arctanh\left( \frac{z}{\sqrt{1+z^2}} \right)$. In the limit $a\to\infty$ (i.e.\ $R_{\max}\to\infty$) one recovers $|I|_T\to(a/c)\arctanh(|J|/c)$.}

\subsubsection{Half-sweep for reflecting geodesics ($|J|>c$)}
For reflecting geodesics, the point of closest approach is the turning point $u_\star=u_*$, given by the condition \eqref{turning-point-condition}. Therefore, the half-sweep integral takes the form
\begin{equation}\label{half-sweep-bouncing}
    |I|_R=\int_{|J|}^{R_{\max}} \frac{|J|\,a\,dR}{R\sqrt{R^2-J^2}\sqrt{R^2-c^2}} \,.
\end{equation}

The integrand again has an integrable singularity at the lower limit, $R=|J|$. To regularize it, we introduce the substitution \mbox{$R=|J|\cosh y$}, obtaining
\begin{equation}
    |I|_R=a\int_0^{y_R}\frac{dy}{\cosh y\,\sqrt{J^2\cosh^2 y-c^2}}\,.
\end{equation}

where $\cosh y_R = R_{\max}/|J|$. Likewise, we introduce the substitution $t=\tanh y$, with $t_R\equiv\tanh y_R=\sqrt{R_{\max}^2-J^2}/R_{\max}$, obtaining the half-sweep
\begin{equation}\label{half-sweep-bouncing-result}
    |I|_R=a \int_0^{t_R}\frac{dt}{\sqrt{J^2-c^2+c^2t^2}} = \frac{a}{c}\arctanh\left(\frac{c\sqrt{a^2+c^2-J^2}}{a|J|}\right)\,.
\end{equation}

The total angular sweep for both types of geodesics is
\begin{align}
    &{|\Delta v_{\mathrm{tot}}|}_T=2\,\frac{a}{c}\arctanh\left(\frac{a|J|}{c\sqrt{a^2+c^2-J^2}}\right) \,, \label{total-half-sweep-result-T}\\
    &{|\Delta v_{\mathrm{tot}}|}_R=2\,\frac{a}{c}\arctanh\left(\frac{c\sqrt{a^2+c^2-J^2}}{a|J|}\right) \label{total-half-sweep-result-R} \,.
\end{align}

The deflection angle \eqref{deflection-angle} then follows from \eqref{theta-in-out-result} and Eqns.~\eqref{total-half-sweep-result-T} and \eqref{total-half-sweep-result-R} as
\begin{align}
    & \delta_T=2\,\frac{a}{c}\arctanh\left(\frac{a|J|}{c\sqrt{a^2+c^2-J^2}}\right)  \, ,\\
    &\delta_R=2\,\arcsin \left(\frac{|J|}{\sqrt{a^2+c^2}}\right)+2\,\frac{a}{c}\arctanh\left(\frac{c\sqrt{a^2+c^2-J^2}}{a|J|}\right)-\pi \,.
\end{align}

\section{Geodesic traversability: a ray limit of reflection and transmission coefficients}

We have a geodesic traversability on a fixed spatial geometry with a throat and two openings. Our spacetime is a product $\mathbb{R}_t\times\Sigma$ with ($\fermi=1$)
\begin{equation}
    ds^2=dt^2-(du^2+R^2(u)dv^2) \ , \qquad R(u)=c\cosh\ur \ .
\end{equation}
The geodesics in this metric are reduced to spatial geodesics on $\Sigma$ (up to a trivial reparametrization in $t$).

From the geodesic equation for the surface-of-revolution spatial metric $dl^2=du^2+R^2(u) \ dv^2$, we have first integrals
\begin{equation}\label{1st-ints-motion}
    \dot{v}=\frac{J}{R^2(u)} \ ,\qquad \dot{u}^2=1-\frac{J^2}{R^2(u)} \ .
\end{equation}
The first relation defines the conserved angular momentum $J$, whereas the second equation can be recast as a one-dimensional Newtonian conservation law, $\frac{1}{2}\dot{u}^2 + V_{\text{eff}}(u) = E$, with the effective potential given by
\begin{equation}
    V_{\text{eff}}(u) = \frac{1}{2}\frac{J^2}{R^2(u)} \,.
\end{equation}
By directly matching this structural form with the geometric constraint $\dot{u}^2 + J^2/R^2(u) = 1$, the total mechanical energy of the trajectories is identically fixed to the constant value $E = 1/2$. Consequently, the spatial geodesic analysis provides a consistent short-wavelength ray limit of the underlying wave equation.

In \secref{geodesics-HYP-classification-section}, we learned that geodesics are classified by the absolute value of their angular momentum. If we define the critical value of angular momentum as
\begin{equation}
    J_c=c=R_{\text{min}} \ ,
\end{equation}
the geodesic rays classify into:
\begin{align}
    &|J|<J_c \qquad (\text{traversing rays}) \ , \\
    &|J|>J_c \qquad (\text{reflecting rays}) \ , \\
    &|J|= J_c \qquad (\text{critical orbit}) \ .
 \end{align}

The near-critical case $|J|\approx J_c$ corresponds to geodesics that wind many times around the throat before escaping. From the unit arc length parametrization of geodesics, $(\frac{du}{dl})^2+R^2(u)(\frac{dv}{dl})^2=1$, we find
\begin{equation}
    \frac{du}{dl}=\sqrt{1-\frac{J^2}{R^2(u)}} \ .
\end{equation}
For light-like quasiparticles, we have $ds^2=dt^2-dl^2=0$, thus $dt=dl$. We obtain that the time spent by the geodesic on its path is the same as the lab time, given by
\begin{equation}\label{lab_time}
    t_{\text{lab}}=\int_{u_i}^{u_f}\frac{du}{\sqrt{1-\frac{J^2}{R^2(u)}}} \ .
\end{equation}

We are interested in how this integral behaves near the throat $u=0$ for the geodesics with the critical value of angular momentum. The function $R(u)= c \cosh\ur$ near the throat:
\begin{equation}
    R(u)\approx c \ \left( 1+\frac{u^2}{2 a^2}\right) \ .
\end{equation}
Considering the critical geodesics with $J_c=c$, we obtain that the integrand behaves as $\sim\frac{1}{|u|}$, which leads to a logarithmically divergent integral \eqref{lab_time}.

The geodesics with the value of angular momentum only close to critical, that is $|J|\approx J_c$, spend a very long time around the throat of the geometry until they eventually escape. At the classical level, this provides the foundation for the appearance of resonances in the corresponding quantum problem.

We can introduce a classical transmission/reflection function.
\begin{equation}
    T_{\text{cl}}(J)=\Theta(c-|J|) \ , \qquad R_{\text{cl}}(J)=\Theta(|J|-c) \ ,
\end{equation}
where $\Theta$ is the Heaviside step function. Classically, we have a strict step, in the quantum mechanical case, these coefficients are smeared and take continuous values between 0 and 1.

\section{Conclusions and future prospects}\label{Summary}

The structure of the BTZ metric provides a geometric rationale for evaluating the Euclidean embeddings of the BEL, ELL, and HYP pseudospheres in $\mathbb{R}^3$, as the spatial section of the BTZ spacetime maps onto specific pseudospherical profiles depending on the mass parameter $M$. This analysis indicates that the hyperbolic pseudosphere (HYP) represents a suitable candidate for experimental realizations in carbon-based systems, owing to its open throat geometry and the finite coordinate boundaries of its embedding domain along the radial coordinate $u$.

The mapping between the HYP geometry and the corresponding gravitational analogue shows that the ultrastatic graphene metric \eqref{ultrastatic-graphene-metric} reproduces the spatial profile of a traversable Morris-Thorne wormhole spacetime with a vanishing redshift function, $\Phi=0$. In standard general relativity, the functional form of the wormhole shape function $b(r)$ is determined dynamically by the stress-energy tensor through the Einstein field equations. This constitutes a structural boundary for the graphene analogue, as the condensed matter system does not incorporate gravitational dynamics or backreaction phenomena; the carbon membrane serves as a fixed background geometry that does not deform in response to the quasiparticle current.

Nevertheless, the system accommodates the exact kinematics of spinorial light rays, represented by the continuous massless Dirac fermions near the charge neutrality point. By solving the spatial geodesic relations on the hyperbolic manifold, the trajectories are classified into traversing and reflecting rays based on the magnitude of their conserved angular momentum $|J|$ relative to the throat radius.

In this geometric ray limit, the throat of the HYP establishes independent transmission and reflection channels, where the critical angular momentum $J_c = c$ defines a sharp classical transmission threshold. Near this threshold ($|J| \approx J_c$), the trajectories undergo a logarithmic divergence in their angular evolution, causing the carriers to execute multiple windings and experience a significant kinematic delay within the throat region. This classical orbiting behavior provides the semiclassical framework necessary to predict quantum-mechanical corrections, where these long-lived states are expected to govern the emergence of sharp scattering resonances within the multichannel Landauer-B\"uttiker transport description.

\section*{Acknowledgements}

We thank Raffaele Agostino, Pavel Krtou\v{s}, Andrea Legramandi and Petr Luke\v{s} for valuable discussions. The financial support from Charles University Research Center (UNCE 24/SCI/016) is gladly acknowledged. A.I. also thanks the VIS programme of Regione Calabria for supporting a visiting professorship to University of Calabria, where part of this work was carried out.

%

\appendix

\section{Pseudospheres}\label{appendixPseudospheres}

\begin{figure}[h]
    \centering
    \includegraphics[width=0.7\linewidth]{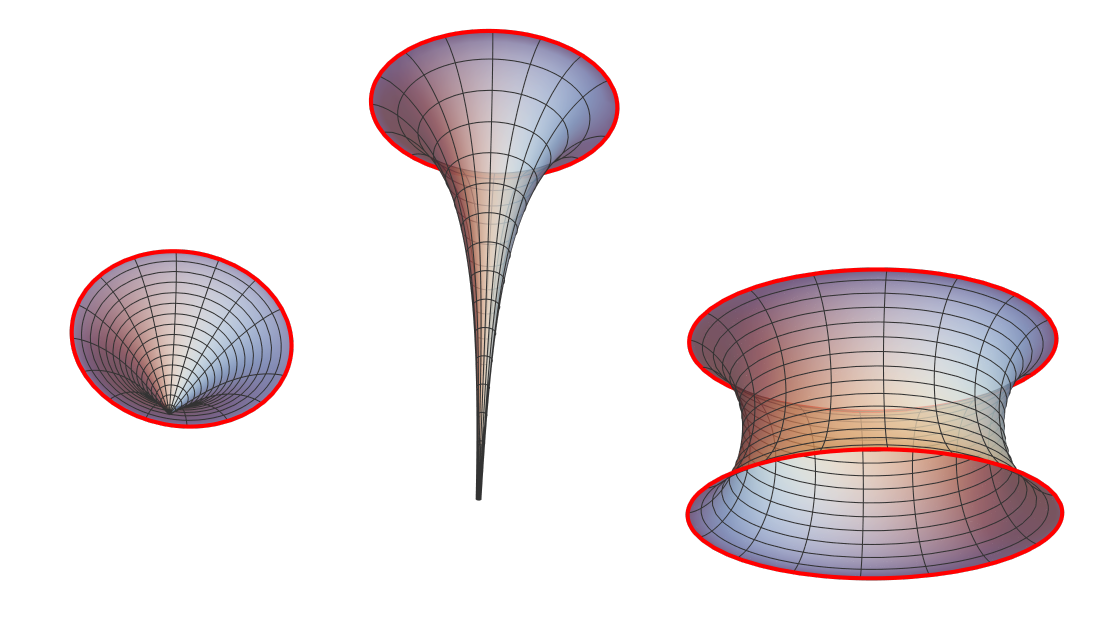}
\caption{The three pseudospheres with their boundaries (in red), called in \cite{iorio2014} ``Hilbert horizons''.}
    \label{fig:3 pseudosp ELLBELHYP bound}
\end{figure}

A pseudosphere in Lobachevsky geometry is the counterpart of the sphere $S^2$ in Euclidean geometry. In other words, whereas for the sphere of radius $r$ the Gaussian curvature $\mathcal{K}=+1/r^2$ is constant and positive, a pseudosphere is a two-dimensional surface for which $\mathcal{K}=-1/r^2$ is constant and \textit{negative}. Actually, in contrast to $S^2$, which may be considered as unique (without going into the details of different coordinate charts and mappings), surfaces of constant negative curvature $\mathcal{K}$ are infinitely many in number \cite{ovchinnikov,mclachlan} (see also \cite{eisenhart1909treatise}, and, e.g., the discussion in \cite{iorio2015}). Among them, only three are proper pseudospheres, namely those that are also surfaces of revolution.

Surfaces of revolution in $\mathbb{R}^3$ are obtained by rotating a curve $\gamma:[A,B]\rightarrow\mathbb{R}^2$, lying in the $(x,z)$-plane, through an angle of $2\pi$ about the $z$-axis (the conventional choice). Such surfaces can be parameterized by the coordinates $(u,v)$, where $v\in [0,2\pi]$ is the azimuthal angle and $u\in[A,B]$ is the parameter along the meridian of the surface (see e.g. \cite{eisenhart1909treatise}).
	
The embedding of the surface of revolution into $\mathbb{R}^3$ is given by
\begin{equation}\label{emb}
x(u,v)=R(u)\cos v, \ \ y(u,v)=R(u)\sin v, \ \ z(u)=\pm\int^{u}\sqrt{1-{R'}^2} d\bar{u},
\end{equation}
where $R'= dR/d\bar{u}$. The profile function $R(u)$ represents the radial distance of the generating curve $\gamma$ from the $z$-axis. Moreover, $R(u)$, together with its domain and range, uniquely determines the surface of revolution up to the choice of orientation. Choosing the parametrization such that $E=1$ in the first fundamental form (with $F=0$ and $G=R^2(u)$) implies the constraint
\begin{equation}\label{arc-length-constraint}
    R'^2(u)+z'^2(u)=1 \,.
\end{equation}
The induced metric on the surface of revolution is therefore
\begin{equation}\label{arc-length-parametrization-metric}
dl^2=du^2+R^2(u) \, dv^2 \,,
\end{equation}
where $u$ is the arc-length parameter along the generating curve.

The Gaussian curvature of the surface is given by \cite{eisenhart1909treatise}
\begin{equation}\label{curvature}
\mathcal{K}=-\frac{R''(u)}{R(u)}.
\end{equation}
This is an easy-to-solve differential equation for the profile function $R(u)$
\begin{equation}\label{ODEforR}
R''(u) + \mathcal{K} \, R(u) = 0 \,.
\end{equation}
The general solution for constant negative Gaussian curvature, \mbox{$\mathcal{K} = - 1/a^2$}, is
\begin{equation}\label{GenSolR}
R (u) = c_1 \cosh (u/a) + c_2 \sinh (u/a) \,,
\end{equation}
where $c_1$ and $c_2$ are arbitrary real constants. By choosing the integration constants appropriately, we obtain the profile functions corresponding to the three classical pseudospheres \cite{eisenhart1909treatise}
\begin{align}
\text{elliptic (ELL):} \ \ \ \ &R(u)=c \ \sinh(u/a) \,, \label{ELL} \\
\text{Beltrami (BEL):} \ \ \ \ &R(u)=c \ e^{u/a} \,, \label{BEL} \\
\text{hyperbolic (HYP):} \ \ \ \ &R(u)=c \ \cosh(u/a) \,, \label{HYP}
\end{align}
where for the elliptic pseudosphere, the integration constant $c$ is chosen such that $0<c=a\sin\theta<a$, whereas for the rest of the pseudospheres, $c$ is real and positive.

For a real embedding in $\mathbb{R}^3$, the constraint \eqref{arc-length-constraint} requires $z'^2(u)=1-R'^2(u)\ge 0$. This condition determines the admissible domains of the meridional coordinate $u$ for the pseudospheres:
\begin{align}
     \ & u_{ELL}\in[0, a \, {\rm arcsinh}\cot\theta] \,, \label{ELL-u-range}\\
     \ & u_{BEL}\in(-\infty, a \, \ln (a/c)] \,, \label{BEL-u-range} \\
     \ & u_{HYP}\in\left[ -a \, {\rm arccosh} \sqrt{1+\frac{a^2}{c^2}},  a \, {\rm arccosh}\sqrt{1+\frac{a^2}{c^2}} \right] \,. \label{HYP-u-range}
\end{align}
Therefore, the corresponding ranges of the profile functions are
\begin{align}
 \quad & R_{ELL}(u)\in[0,a\cos\theta] \,, \label{ELL_range} \\
 \quad & R_{BEL}(u)\in(0,a] \,, \label{BEL_range} \\
 \quad & R_{HYP}(u)\in[c,\sqrt{a^2+c^2}] \,. \label{HYP_range}
\end{align}
Notice that the HYP has the symmetric domain \eqref{HYP-u-range}, meaning that the maximum of the profile function $R_{HYP}(u)$ is reached at both ends of the interval. The minimal value $R_{HYP}(0)=c$ of the profile function occurs in the middle of the domain, \mbox{$u_{HYP}=0$}.

The three profile functions \eqref{ELL}--\eqref{HYP} describe three distinct geometries according to the way the surfaces approach the $z$-axis, as follows from the domains \eqref{ELL-u-range}--\eqref{HYP-u-range}. The hyperbolic pseudosphere has a minimal circle at $u_{HYP}=0$, where $R_{HYP}=c$; the Beltrami pseudosphere approaches the $z$-axis only asymptotically at $u_{BEL}\rightarrow -\infty$, forming an infinitely long cusp; and the elliptic pseudosphere reaches the $z$-axis at the finite value \mbox{$u_{ELL}=0$}.

\section{The Necessity of the Dual Requirement for Surface Gravity}\label{appendixConfFlatness}

A central conceptual question within the Weyl-symmetry framework addresses whether conformal flatness is strictly required to establish a conformal Killing horizon (CKH), or if it is sufficient for a physical metric to be merely conformal to a spacetime endowed with a Killing vector field.

Geometrically, the second condition is self-sufficient: any metric related to a Killing background by $g_{\mu\nu}^B = \varphi^2 g_{\mu\nu}^A$ redefines the original Killing horizon (KH) as a CKH. However, within the physical constraints imposed by the flat temporal gauge of Eq.~\eqref{mainmetric}, conformal flatness constitutes an indispensable prerequisite.

If an arbitrary conformal factor $\varphi^2$ is applied to a generic black hole or cosmological metric, the resulting manifold does not, in general, preserve the synchronous conditions $g_{00} = 1$ and $g_{0i} = 0$. The uniformity of the laboratory time evolution would be compromised. Consequently, conformal flatness represents the mathematical condition that simultaneously satisfies three distinct operational constraints:
\begin{itemize}
    \item Preservation of a flat temporal evolution ($g_{00} = 1$, $g_{0i} = 0$), matching laboratory clocks.
    \item Generation of a non-trivial, curved spatial background for the carbon membrane ($g_{ij} = g_{\text{pseudosphere}}$).
    \item Exploitation of conformal triviality to project the background curvature into a global spatial factor $\varphi^{-2}$ multiplying the local density of states (LDOS).
\end{itemize}

This mechanism provides a consistent physical mapping of the Tolman relation within the analog framework. If the electronic spectrum is evaluated relative to the laboratory time $t$ (as dictated by a standard scanning tunneling microscopy setup), the temperature parameter in the thermal distribution\footnote{In odd dimensions, the phenomenon of statistical inversion between Bose-Einstein and Fermi-Dirac distributions is predicted, known as Takagi's statistics inversion \cite{Takagi:1986kn}. Consequently, the thermal distribution at the Unruh-Hawking temperature follows Bose-Einstein statistics despite the fermionic nature of the quasiparticles.} remains the unrescaled horizon temperature ($T_U$), while the spatial geometry modulates the phase-space volume via the multiplicative factor $\varphi^{-2}$. Conversely, if the spectrum is expressed in terms of the local proper time $\tau$ (where $d\tau = \varphi dt$), the spectral amplitude remains unscaled, while the temperature parameter absorbs the spatial modulation, yielding the local physical profile:
\begin{equation}
    T(u) = \frac{T_U}{\varphi(u)} = \frac{\hbar v_F}{2\pi k_B} \frac{\ell}{a^2} e^{u/a} \,.
\end{equation}
Without enforcing conformal flatness at the onset of the ansatz, the simultaneous requirement of a flat temporal gauge and a non-trivial spatial curvature would either restrict the manifold to a flat Minkowski geometry or eliminate the conformal symmetries necessary to sustain a coherent thermodynamic interpretation of the analog horizon.

\section{Geodesics for the ELL}
Similarly as it was done in \secref{geodesics-HYP-section} for the geodesics of the HYP, we can integrate the Eq.~\eqref{geodesics-integral} to obtain the geodesics for the ELL defined by the profile function $R(u)=c\sinh\ua$.

Substituting the profile function $R(u)=c\sinh\ua$ to the Eq.~\eqref{geodesics-integral-2}, and choosing the positive sign for the geodesics, we obtain an indefinite integral:
\begin{equation}
    I(u)=\int\frac{J\,du}{c\sinh\ua\sqrt{c^2\sinh^2\ua-J^2}}\,.
\end{equation}
Introducing the substitution $t=\coth\ua$, we obtain
\begin{equation}\label{geodescics-indefinite-integral-ELL}
    I(u)=-\frac{a}{c}\int\frac{J\,dt}{\sqrt{c^2+J^2-J^2 t^2}} \,.
\end{equation}

The profile function of the ELL is bounded,
\begin{equation}
    R(u) \in[0, a\cos\theta], \quad c=a\sin\theta\,.
\end{equation}
Therefore, the angular momentum of a geodesic is restricted to the range
\begin{equation}
    0<|J|\le a\cos\theta\,.
\end{equation}
Since every allowed value of $|J|$ lies within the range of the profile function, the turning point condition $R(u_*)=|J|$ always has a solution. Therefore, every geodesic on the ELL possesses a turning point. Hence, there is only one type of geodesics (the returning geodesics), which have a turning point at
\begin{equation}
    u_*=a\arcsinh(|J|/c)\,,
\end{equation}
and for all physical values of the angular momentum $|J|>0$, do not reach the tip of the ELL located at $u=0$.

Evaluating the antiderivative \eqref{geodescics-indefinite-integral-ELL}, we obtain the geodesics on the ELL:
\begin{equation}
    v(u)=-\frac{a}{c}\arcsin \left( \frac{J\coth\ua}{\sqrt{c^2+J^2}} \right)\,, \quad |J|\in(0,a\cos\theta]\,.
\end{equation}

\begin{figure}
    \centering
    \includegraphics[width=0.6\linewidth]{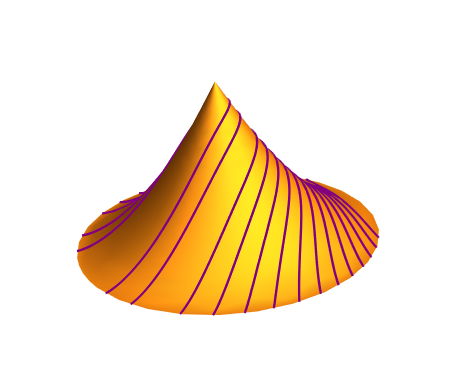}
    \caption{Geodesics for the ELL.}
    \label{ELL-geodesics-figure}
\end{figure}

The geodesics for ELL are plotted in Fig.\ref{ELL-geodesics-figure}.

\section{Semiclassical WKB Expansion and Eikonal Decoupling}
\label{app:WKB}

We evaluate the massless Dirac equation for fermions within the continuous effective theory of graphene near the Dirac points
\begin{equation}
    i \hbar \gamma^\mu(x) \nabla_\mu \psi(x) = 0,
    \label{eq:app_dirac_init}
\end{equation}
where $\gamma^\mu(x) = e^\mu_a(x) \gamma^a$ are the coordinate-dependent Dirac matrices satisfying the Clifford algebra $\{\gamma^\mu(x), \gamma^\nu(x)\} = 2g^{\mu\nu}(x)\mathbb{I}$. The total covariant derivative acting on the spinor field incorporates the spin connection, $\nabla_\mu = \partial_\mu + \Omega_\mu$, which encompasses both the torsion-free Levi-Civita connection $\Omega_\mu^{\text{LC}}$ and any potential contorsion contributions $K_\mu$ arising from lattice torsion.

The physical regime under consideration is governed by the hierarchy of scales $\ell \ll \lambda_e \ll a,c$, where $\ell$ is the lattice spacing and $a,c$ represent the macroscopic parameters of the hyperbolic throat. This hierarchy allows us to introduce a small, dimensionless eikonal parameter:
\begin{equation}
    \epsilon \equiv \frac{\lambda_e}{R_{\text{macro}}} \ll 1.
\end{equation}
The parameter $\epsilon$ operates as a classical geometric counting variable. We look for asymptotic solutions to Eq.~\eqref{eq:app_dirac_init} via the eikonal WKB ansatz
\begin{equation}
    \psi(x) = \chi(x) e^{\frac{i}{\epsilon} S(x)} = \left( \chi_0(x) + \epsilon \chi_1(x) + \mathcal{O}(\epsilon^2) \right) e^{\frac{i}{\epsilon} S(x)},
    \label{eq:app_wkb_ansatz}
\end{equation}
where $S(x)$ is the real scalar action and $\chi_n(x)$ are the spinor amplitudes. This is an adaptation to our emergent set-up of the standard expansion in powers of $\hbar$  \cite{rubinow1963}.

Indeed, an important conceptual distinction must be highlighted regarding the implementation of the expansion within the effective continuum theory of graphene. In standard quantum-mechanical formulations of the WKB method \cite{rubinow1963}, Planck's constant serves as the formal expansion parameter, with the classical limit recovered by evaluating the asymptotic sequence as $\hbar \to 0$. In an emergent condensed matter system, however, $\hbar$ governs the fundamental microphysical scales of the underlying carbon lattice and regulates the definition of the effective Fermi velocity $v_F$. Consequently, $\hbar$ operates as an unalterable coupling constant embedded within the structural equations of the emergent field theory.

To establish a mathematically consistent short-wavelength limit without modifying the invariant role of $\hbar$, we implement an adaptation of the Rubinow-Keller technique tailored to emergent topological landscapes. The asymptotic tracking parameter is defined as a purely geometric, dimensionless ratio $\epsilon \equiv \lambda_e / R_{\text{macro}} \ll 1$, where the excitation wavelength $\lambda_e$ is small compared to the macroscopic geometric scale of the hyperbolic throat. This formulation isolates the short-wavelength ray limit within the laboratory frame, ensuring that the continuous phase $S(x)$ and the corresponding trajectories decouple from the spinorial degrees of freedom at the leading order $\mathcal{O}(1/\epsilon)$, in strict accordance with the classical ray limit of the underlying wave description. Let us see how this comes about.

Substituting Eq.~\eqref{eq:app_wkb_ansatz} into the Dirac equation \eqref{eq:app_dirac_init} and grouping the terms in powers of $\epsilon$ yields:
\begin{equation}
    \frac{1}{\epsilon} \left[ \gamma^\mu(x) \partial_\mu S \right] \chi_0 + \left[ \gamma^\mu(x) \partial_\mu S \chi_1 - i \gamma^\mu(x) \left( \partial_\mu + \Omega_\mu^{\text{LC}} + K_\mu \right) \chi_0 \right] + \mathcal{O}(\epsilon) = 0.
    \label{eq:app_expanded}
\end{equation}

At the leading eikonal order $\mathcal{O}(1/\epsilon)$, Eq.~\eqref{eq:app_expanded} restricts to the localized homogeneous linear system:
\begin{equation}
    \left[ \gamma^\mu(x) \partial_\mu S(x) \right] \chi_0(x) = 0.
    \label{eq:app_leading_matrix}
\end{equation}
To sustain a non-zero propagating electronic current ($\chi_0 \neq 0$), the matrix operator $M \equiv \gamma^\mu \partial_\mu S$ must be non-invertible, requiring $\det M = 0$. Squaring the operator and exploiting the anticommutation relations of the Clifford algebra leads directly to the scalar, spin-independent Hamilton-Jacobi equation for light-like rays:
\begin{equation}
    g^{\mu\nu}(x) \partial_\mu S \partial_\nu S = 0.
    \label{eq:app_eikonal_final}
\end{equation}

Crucially, the total spin connection, including both the Levi-Civita term $\Omega_\mu^{\text{LC}}$ and the contorsion $K_\mu$, is entirely confined to the sub-leading order $\mathcal{O}(1)$ of the expansion. In the strict classical ray limit ($\epsilon \to 0$), these spinorial and torsional degrees of freedom decouple completely from the trajectory determination. The phase $S(x)$ and the corresponding trajectories are governed exclusively by the smooth background metric $g_{\mu\nu}$ through Eq.~\eqref{eq:app_eikonal_final}, fully validating the treatment of the transport problem via the uncoupled, pure metric geodesics of the hyperbolic carbon throat.

\bibliographystyle{unsrt}

\end{document}